\pdfoutput=1
\documentclass[aps,prb,amsmath,amssymb,floatfix,twocolumn,superscriptaddress,nofootinbib,tighten]{revtex4-2}
\usepackage{authoraftertitle}
\usepackage{multirow}
\usepackage{subfigure}
\usepackage{color}
\usepackage{mathrsfs}
\usepackage{hyperref}
\usepackage{physics}
\usepackage[normalem]{ulem}
\usepackage{bm}

\usepackage{amssymb}   
\usepackage{amsmath}
\renewcommand\vec[1]{\ensuremath\boldsymbol{#1}} 

\usepackage{lipsum}
\usepackage{float}
\usepackage{amsfonts, relsize, color}
\usepackage{pifont}
\usepackage{graphics}
\usepackage{graphicx}
\usepackage{subfigure}
\usepackage{color}
\usepackage{comment}

\usepackage{xcolor}
\hypersetup{
	colorlinks,
	linkcolor={red!50!black},
	citecolor={green!50!black},
	urlcolor={blue!50!black}
}
\usepackage{blindtext}

\newcommand{\be}{\begin{equation}}
\newcommand{\ee}{\end{equation}}

\newcommand{\bea}{\begin{eqnarray}}
\newcommand{\eea}{\end{eqnarray}}

\newcommand{\p}{\partial}

\newcommand{\lp}{\left(}
\newcommand{\rp}{\right)}

\renewcommand{\vec}[1]{{\boldsymbol #1}}

\begin{document}

\title{Signatures of electronic ordering in transport in graphene flat bands}

\author{Archisman Panigrahi}\email{archi137@mit.edu}
\affiliation{Department of Physics, Massachusetts Institute of Technology, Cambridge, MA 02139, USA.}
\author{Leonid Levitov} 
\affiliation{Department of Physics, Massachusetts Institute of Technology, Cambridge, MA 02139, USA.}

\date{\today} 

\begin{abstract}
{ 
Recently, a wide family of electronic orders was unveiled in graphene flat bands,  
such as spin- and valley-polarized phases as well as nematic momentum-polarized phases, 
stabilized by 
exchange interactions via a generalized Stoner mechanism. 
Momentum polarization involves orbital degrees of freedom and is therefore expected to impact resistivity in a way which is uniquely sensitive to the ordering type. 
Under pocket polarization, carrier distribution shifts in $k$ space and samples the band 
mass in regions defined by the displaced momentum distribution.
This makes transport coefficients sensitive to pocket polarization, resulting in the ohmic resistivity decreasing with temperature. 
In addition, it leads to current switching and hysteresis under strong $E$ field. 
Being robust in the presence of electron-phonon scattering, this behavior can serve as a telltale sign of pocket
polarization order. The fast timescale and low dissipation of the switching cycle may be advantageous for highly
applicable memory-dependent resistors, i.e., memristors.
}
\end{abstract}

\maketitle

\section{Introduction}
Recent experiments  
reported on a cascade of strongly-correlated phases in moir\'e graphene multilayers \cite{Cao1, Cao2, Sharpe, SerlinAnomalousHall, Tschirhart, Kerelsky, Choi2019, Jiang, CaoNematic, SaitoPomeranchuk, Zondiner, Rozen, Choi2021, Pierce, Zhang-Li-2024, LuFractionalHall} as well as their non-moir\'e counterparts \cite{Lee,Zhou1, Zhou2, Barrera, Seiler, ChenAnomalousHall, Li1, ZhangSuperconductivity2023}, where the electronic properties can be tuned by altering the external displacement field ($D$) and the electron density ($n$).
Moir\'e graphene hosts flat minibands with high density of states, where the effects of electronic interactions become significant \cite{Lopes_dos_Santos,Mele,SuarezMorell,Bistritzer, Guinea, Isobe, Kozii, Xu, Lin, Kang, Xie, Ochi, Seo, Wu1, Wu2, BultinckAnomalousHall, FractionalTheoryZhihuanDong, ZhouFractional,FractionalTheoryJunkaiDong, Dodaro, Liu, BhowmikReview2023, Fischer}. Likewise, in non-moir\'e stacked graphene the displacement field opens up a gap and the resulting bands become remarkably flat near their extrema, giving rise to similar interaction-driven ordered phases \cite{Castro,Castro2,Stauber,Throckmorton,Min,Nandkishore,CvetkovicBLG,Jung,Chou,Stauber,Gorbar,Zhang, Kharitonov, Ghazaryan, ZhiyuIsospinMomentum, Koshino,FZhang, McCann}. To minimize the exchange energy, the electrons spontaneously break the combined SU(4) symmetries in the spin-valley degrees of freedom, giving rise to valley-polarized and spin-polarized ferromagnetic orders \cite{Sharpe, SerlinAnomalousHall, Tschirhart, Choi2019, Jiang, CaoNematic, SaitoPomeranchuk, Zondiner, Ochi, Seo, Wu1, Wu2, Dodaro, Liu, Stauber, Min, Nandkishore, Jung}.
\begin{figure}[t!]
	\includegraphics[width=0.95\linewidth]{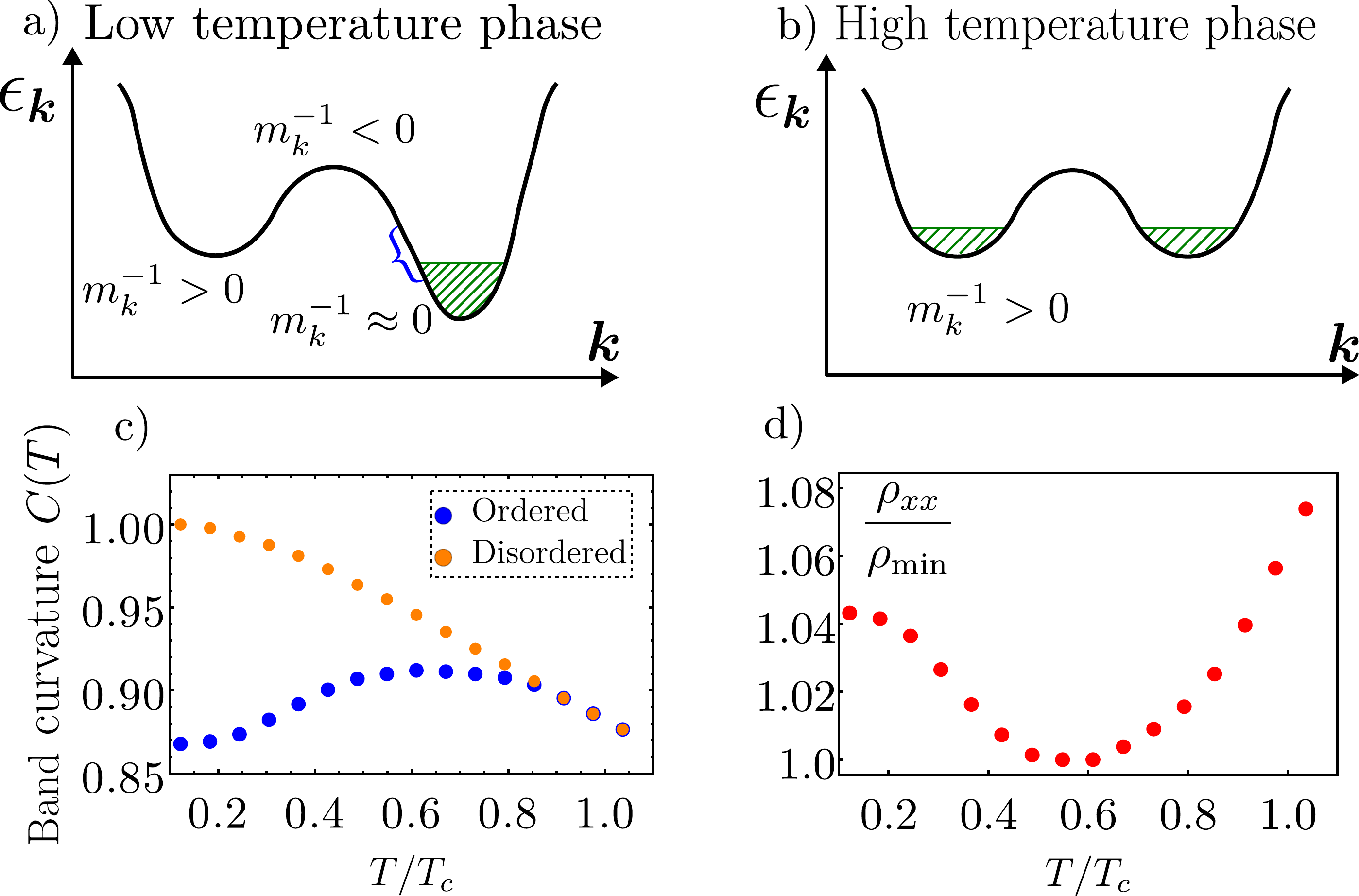}
	\caption{(a), (b) Pocket-polarized and unpolarized states
for 
a double-well 1D model, Eq.\eqref{eq:model1}. Marked are regions in which the band curvature $m_k^{-1}=\partial^2\varepsilon_k/\partial k^2$ is positive, negative, and close to zero. 
	The pocket-polarized carrier distribution shifts in $k$ space and samples the band curvature, Eq.\eqref{Eq:average-curvature}, in regions within the displaced momentum distribution. This makes transport coefficients sensitive to pocket polarization, resulting in ohmic resistance decreasing with temperature (see text). 
		(c) Temperature dependence of the occupancy-weighted band curvature $C(T)$, Eq.\eqref{Eq:average-curvature}, for the ordered states (blue) and the metastable disordered states (orange). 
		(d) 
		The resulting resistivity temperature dependence is nonmonotonic. Initially, resistivity drops 
		due to the increase in the band curvature of the ordered phase. At higher temperature the phonon scattering grows, and the resistivity increases linearly. Parameter values used are given in Sec.\ref{sec:microscopic-model}.
	}~\label{fig:1D}
\end{figure}
Electronic ordering of a different kind can arise in non-moir\'e graphene multilayers when electrons partially fill the pockets induced by trigonal warping of carrier bands 
near the $\vec{K}$ and $\vec{K'}$ points. Such `carrier flocking' in momentum space, driven by electron exchange interaction and described by Stoner-type instability, produces nematic momentum-polarized orders where only some of the pockets remain populated \cite{Seiler, Li1, ZhiyuIsospinMomentum}. 
Upon temperature increasing, the ordering disappears through what is currently believed to be a continuous phase transition. 

Several different techniques have been employed to probe electronic orders at low temperatures. Phase boundaries between different ordered phases can be probed by electronic compressibility measurements which capture the change in density of states at the Fermi surface, across the phase transition\cite{Zhou1,Barrera,Seiler}. The low-temperature momentum-polarized, spin-polarized, and valley-polarized orders can be identified by quantum oscillations measurements\cite{Zhou2}, which reveal the spin/valley degeneracy and the Fermi surface geometry.

In comparison, the signatures of ordering in the temperature dependence of transport coefficients have received relatively little attention.
The only transport signature explored so far, which distinguishes symmetry-broken phases, was the anomalous Hall effect observed in the valley-polarized phase at low temperature, 
which disappear at elevated temperatures\cite{ChenAnomalousHall, SerlinAnomalousHall, BultinckAnomalousHall}. In that, the $\vec{K}$ and $\vec{K'}$ valleys, which have opposite signs of Berry curvature, 
give rise to an anomalous Hall response resulting from broken time-reversal symmetry.

Motivated by recent findings \cite{AFYPrivateCommunication}, in this work we consider the 
momentum-polarized order perturbed by a DC electric current, both in the Ohmic regime (linear field-current response), and 
the non-Ohmic (nonlinear response) regime.  Instead of the the Berry curvature, we focus on the band dispersion and its dependence on the pocket-polarization ordering.
As we will see, under 
such ordering, carrier distribution shifts in $k$ space and samples the band curvature in regions defined by the displaced 
Fermi sea. This behavior makes transport coefficients sensitive to pocket polarization  
and leads to a unique signature of ordering---the negative temperature dependence of resistivity below the order-disorder transition. Namely, rising temperatures, despite increasing electronic disorder, drive the system into a more conducting state. This 
 behavior originates from an interplay between exchange interactions that make carriers aggregate in Fermi pockets and thermal activation that excites carriers out of the Fermi pockets.
\begin{figure}[t!]
\includegraphics[width=0.85\linewidth]{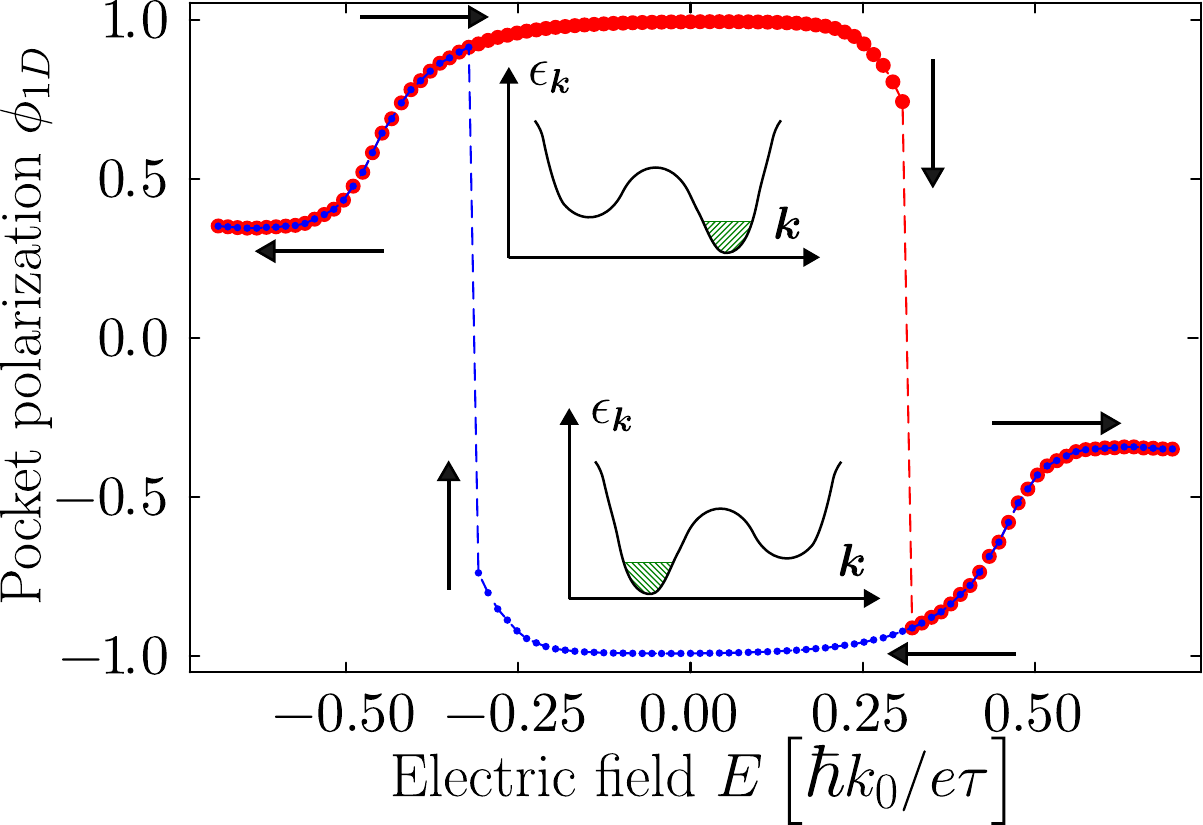}
	\caption{
	Pocket polarization switching and history dependence of the dynamics under the action of an electric field for the model illustrated in Fig.\ref{fig:1D}. 
As discussed in Sec.\ref{sec:switching}, the initial state is the equilibrium state polarized in the right (left) pocket for the red (blue) curve. For each value of electric field $E$, we self-consistently determine the steady state, and plot the pocket polarization $\phi_{1D} = (n_R-n_L)/(n_R+n_L)$ against the (dimensionless) electric field. The right and left pocket-polarized steady states are pictured inside the hysteresis loop. For the right pocket-polarized initial state (red curve), for positive electric fields, the Fermi sea shifts to the $-\hat{k}_x$ direction, and the distribution switches to the left pocket at a finite electric field. No such switching occurs when the electric field is swept in the other direction. The opposite effect occurs for the left-pocket polarized initial state (blue curve). The black arrows show the direction of switching of the steady state when the electric field is swept. 
	}~\label{fig:E_field_switching}
\end{figure}

Furthermore, under the application of a strong electric field, the momentum-polarized phase can exhibit history dependence and switching of pocket polarization. This behavior is illustrated in Fig.\ref{fig:E_field_switching} which displays the results of a microscopic analysis developed in Sec.\ref{sec:switching}. The resulting current switching 
may serve as an experimental test 
for discerning the proposed mechanism of negative temperature dependence of resistivity from alternative mechanisms. 
Similar switching
behavior has been observed in recent experiments \cite{SerlinAnomalousHall, TschirhartSwitching2023, Zhang-Li-2024}, in twisted bilayer graphene, transition metal dichalcogenides, and twisted trilayer graphene. It has been studied theoretically for magic-angle twisted bilayer graphene \cite{SuSwitching2020}.

The switching mechanism in these systems does not involve any movement of impurities, nor does it involve any movement of electrons in the real space. Instead, it is driven by carrier distribution switching between different pockets in momentum space. Consequently, the switching timescale will be very fast, which opens up the possibility of using these systems as memory-dependent resistors, i.e., memristors \cite{ChuaMemristor1971}. Moreover, since the switching is observed at relatively small currents (nano-Ampéres) \cite{Zhang-Li-2024}, the dissipation per switching cycle is expected to be small, potentially leading to highly efficient memristors.

\section{Transport anomalies due to pocket polarization ordering}\label{sec:transport-anomalies}

The system in question, where a phase transition alters the geometry and symmetry of the Fermi sea, features band dispersion which is uniquely sensitive to carrier momentum distribution.
Namely, the phase transition changes the band curvature 
$m_{k_ik_j}^{-1}=\p^2 \varepsilon_{\vec k}/\p k_i \p k_j$ defined by the part of the band occupied by carriers,
\begin{equation}\label{Eq:average-curvature}
C(T)=\left\langle{\frac{\partial^2 \varepsilon_{\vec{k}}}{\partial k_x^2}}\right\rangle_T = \frac{1}{n} \sum_{\vec k, \sigma} \frac{\partial^2 \varepsilon_{\vec{k},\sigma}}{\partial k_x^2} f_{\vec{k},\sigma}
,
\end{equation}
where $f_{\vec{k},\sigma}$ is momentum distribution and $\varepsilon_{\vec{k}}$ is band dispersion which in itself depends, via exchange interaction, on $f_{\vec{k},\sigma}$. (Here $n=\sum_{\vec k,\sigma} f_{\vec{k},\sigma}$ is carrier density.) As illustrated in Fig.\ref{fig:1D}(a) and (b), 
different parts of a two-pocket carrier band have curvature of opposite signs and, therefore, give opposite-sign contributions to $C(T)$.

As we will see, the changes in $C(T)$ driven by ordering translate into a characteristic $T$ dependence of the resistivity.
We will outline a mechanism through which the resistivity in the ordered phase at low-temperature can be higher than that of the disordered phase at a higher temperature.
In other words, we argue that the resistivity can decrease with rising temperature, as the ordered phase melts.
Negative temperature dependence, $d\rho/dT<0$, stands in contrast with the sign of temperature dependence typically 
seen in metals due to carrier scattering by lattice vibrations. The low characteristic temperature at which it is expected to occur as well as the sign, make the negative $d\rho/dT$ a telltale sign of pocket polarization ordering.

The mechanism leading to negative $d\rho/dT$ is illustrated in Fig.\ref{fig:1D}.
In Boltzmann transport theory 
\cite{AshcroftMermin} the DC 
ohmic conductivity 
is given by the expression,
\begin{equation}\label{eq:sigma_xx}
		\sigma_{xx}(T) = \sum_{\vec k, \sigma} 
		\frac{e^2 \tau(T)}{\hbar^2} \frac{\partial^2 \varepsilon_{\vec{k},\sigma}}{\partial k_x^2} f_{\vec{k},\sigma} 
		= \frac{n e^2 \tau(T)}{\hbar^2} C(T),
\end{equation}
where $f_{\vec k, \sigma}$ is 
the momentum distribution,
$\sum_{\vec k}$ is a shorthand for $\int \frac{d^2\vec{k}}{(2\pi)^2}$ and, for simplicity, the relaxation timescale $\tau$ is assumed to be momentum independent. Clearly, Eq.\eqref{eq:sigma_xx} implies that between two phases with comparable $\tau$ values, 
the one with greater average band curvature $C(T)$ will have a smaller resistivity. 

The quantity $\varepsilon_{\vec{k},\sigma}$ in Eq.\eqref{eq:sigma_xx} denotes the 
band dispersion 
modified by interactions, and $f_{\vec k, \sigma}$ is the Fermi function for this band dispersion. 
Under mean-field theory developed below, the pocket-asymmetric part of the  momentum distribution $f_{\vec k, \sigma}$ plays the role of the order parameter describing pocket polarization. The energy $\varepsilon_{\vec{k},\sigma}$ depends on 
$f_{\vec{k}, \sigma}$ and the non-interacting dispersion $\varepsilon^0_{\vec{k}}$ as
\begin{align}\label{eq:mean-field-gen-sol}
&\varepsilon_{\vec{k},\sigma} = \varepsilon^0_{\vec{k}} - \sum_{\vec{k}'\neq \vec{k}} V(\vec{k}-\vec{k}') f_{\vec{k}',\sigma},
\end{align}
where $f_{\vec{k}, \sigma}$ is given by a Fermi distribution for the band dispersion modified by interactions,
\begin{align} \label{eq:mean-field-occupation-function}
& f_{\vec{k},\sigma} = \frac{1}{e^{\beta (\varepsilon_{\vec{k},\sigma} - \mu)}+1},
\end{align}
and $V(\vec{k}-\vec{k'})$ 
represents the electronic interactions (see Sec.\ref{sec:microscopic-formalism}). Eqs.\,\eqref{eq:mean-field-gen-sol} and \eqref{eq:mean-field-occupation-function}, solved self-consistently, describe 
carrier momentum distribution in the ground state in the presence of exchange interactions. When interactions are strong enough, the momentum distribution
becomes pocket-asymmetric at low temperature, as illustrated in Figs. \ref{fig:1D} and \ref{fig:E_field_switching}. The occurrence of multiple fixed points for the self-consistent solution of these equations is a signature of a pocket symmetry breaking instability towards pocket polarization order. Upon temperature increasing, the solution undergoes a transition to a disordered pocket-symmetric state.

We now discuss reasons for which the average occupancy-weighted band curvature, Eq.\eqref{Eq:average-curvature},
can be lower in the low temperature phase than that in the corresponding high-temperature disordered phase, thus leading to resistivity decreasing with temperature.
As an illustration, consider the momentum-polarized to momentum-unpolarized transition, focusing, for simplicity, on the pocket-polarized order in a single valley with a single spin species. Bilayer graphene hosts three pockets (or four pockets, at small enough displacement field) induced by trigonal warping near its $\vec{K}$/$\vec{K'}$ points \cite{Seiler2024}. The regions in $k$-space where these pockets merge have a negative band curvature (hole-like), whereas the minima of the pockets have a positive band curvature. At low temperature, the ordered phase with a high enough electron density will feature a band filled upto the neck of a single pocket (Fig.\ref{fig:1D}(a)). The band curvature of this pocket near the chemical potential will be much smaller than that near the pocket bottom. As temperature is increased, the electrons will be thermally excited to (i) high-energy states within the same pocket, which either have a positive band curvature with smaller magnitude, or a curvature with negative sign, and, (ii) the low-energy states of the previously empty pocket, where the band curvature is positive and has relatively large magnitude. Clearly, there will be a competition between the two effects. When the electron density is high enough such that the chemical potential at low temperature is close to the bottom of the empty pocket, the second effect wins due to the high density of states, 
and the occupancy-averaged band curvature increases with temperature.

Experimentally, such an effect may occur on the high density side of the momentum-polarized phase in the $n-D$ phase diagram, $D$ being the displacement field.
The density needs to be such that the Fermi energy of the filled pocket is close to the bottom of the empty pocket, so that due to thermal excitations, electrons will be excited to the bottom of the empty pocket, decreasing the average curvature (of course, there will be a competition between the higher energy states of the partially filled pocket and the bottom states of the empty pocket). To the contrary, when the density is low, the distance between the Fermi energy at the filled pocket and the bottom of the empty pocket may be too large for thermal excitations to excite the electrons to the bottom of the empty pocket.


At temperatures below the Bloch-Gr\"uneisen temperature
\cite{AshcroftMermin,Stormer1990, Hwang-DasSarma2008}
phonons are not yet thermally activated, so the carrier momentum relaxation timescale $\tau(T)$ remains almost temperature independent. At such temperatures, the resistivity will primarily depend on the band curvature. As temperature varies, the resistivity initially decreases with rising temperature as the average band curvature increases due to gradual melting of the ordered phase. Eventually, the resistance will begin to grow with temperature as the phonons become thermally activated. In Sec. \ref{sec:microscopic-model} we establish this effect with two microscopic models that resemble the pockets in the bilayer graphene  bandstructure (see Figs.\ref{fig:1D}(d),\ref{subfig:resistivity-2d}).


It is instructive to compare the resulting temperature dependence, which is fairly strong, to various scenarios discussed in the literature. Phase transitions in metals of first order usually result in resistance  discontinuity at the phase transition but  little $T$ dependence away from it. Phase transitions of second order in general lead to strong order parameter fluctuations at $T$ above and below the transition, and a singularity at the transition. These fluctuations, however,  are of a long-wavelength character, and, as a result, do not produce  a strong $T$-dependence of transport cross-section and resistivity \cite{Fisher-Langer1968}.
In contrast, electronic ordering in graphene flat bands discussed here impacts carrier scattering at large angles, i.e. 
in a wide range of transferred momenta $k$. This makes resistivity sensitive to the ordering type, giving rise to the $T$ dependence discussed below.

We also comment on several other effects due to electron interactions that can potentially result in a negative temperature dependence of resistivity, $d\rho/dT<0$. 
Some years ago, 
ultraclean silicon MOSFETs 
were found to exhibit negative $T$ dependence of resistivity. A number of mechanisms to explain this behavior have been proposed. Those included the Coulomb scattering crossection weakening at a higher temperature\cite{GoldDogolopov}, an effect that leads to a decrease in the disorder scattering rate. 
Other explanations focused on the effects due to the presence of charge traps and
the Altshuler-Aronov-type interaction effects \cite{Maslov1,Maslov2,Zala2001}.

Furthermore, a very different physics resulting in a negative $T$ dependent resistivity was proposed for clean metals, where carrier collisions can result in hydrodynamic transport, wherein electrons undergoing two-body collisions behave as a viscous fluid \cite{Gurzhi1968}. Electron hydrodynamics can naturally lead to resistivity that decreases with temperature. Later, however, it has been argued that the actual behavior is more nuanced \cite{Andreev2011}. Namely, correlated electron systems in a slowly varying disorder potential were predicted to show resistivity with a growing T dependence, an effect arising due to the interplay of heat conduction and thermoelectric effects. More recently, however, hydrodynamic transport in clean electron systems with sharp boundaries was confirmed to result in a negative temperature dependence of resistivity \cite{HGuo2017, tomogrph}. The prediction of hydrodynamic resistivity decreasing with temperature is supported by recent measurements \cite{Roshan2017, Ginzburg2021}.


This behavior must therefore be accounted for in delineating mechanisms leading to a negative $T$ dependence in realistic systems. We note in that regard that there are several qualitative aspects of the pocket-polarization mechanism for negative temperature dependence of resistivity, $d\rho/dT<0$, which make it distinct from the other known mechanisms. One is its lack of sensitivity to the cleanness of the system. This stands in contrast to electron hydrodynamics which requires the electron system to be ultra-clean. Another is that it is tied to temperature at which the pocket polarization phase transition occurs. Lastly, as discussed below, 
pocket polarization order gives rise to a switching behavior of the nonlinear $I$-$V$ dependence which coexists with the negative temperature dependence of resistivity. These properties occurring together will provide clear signatures of transport anomalies due to pocket polarization instability.

\begin{figure}[t!]
\subfigure[]{\includegraphics[width=0.5\linewidth]{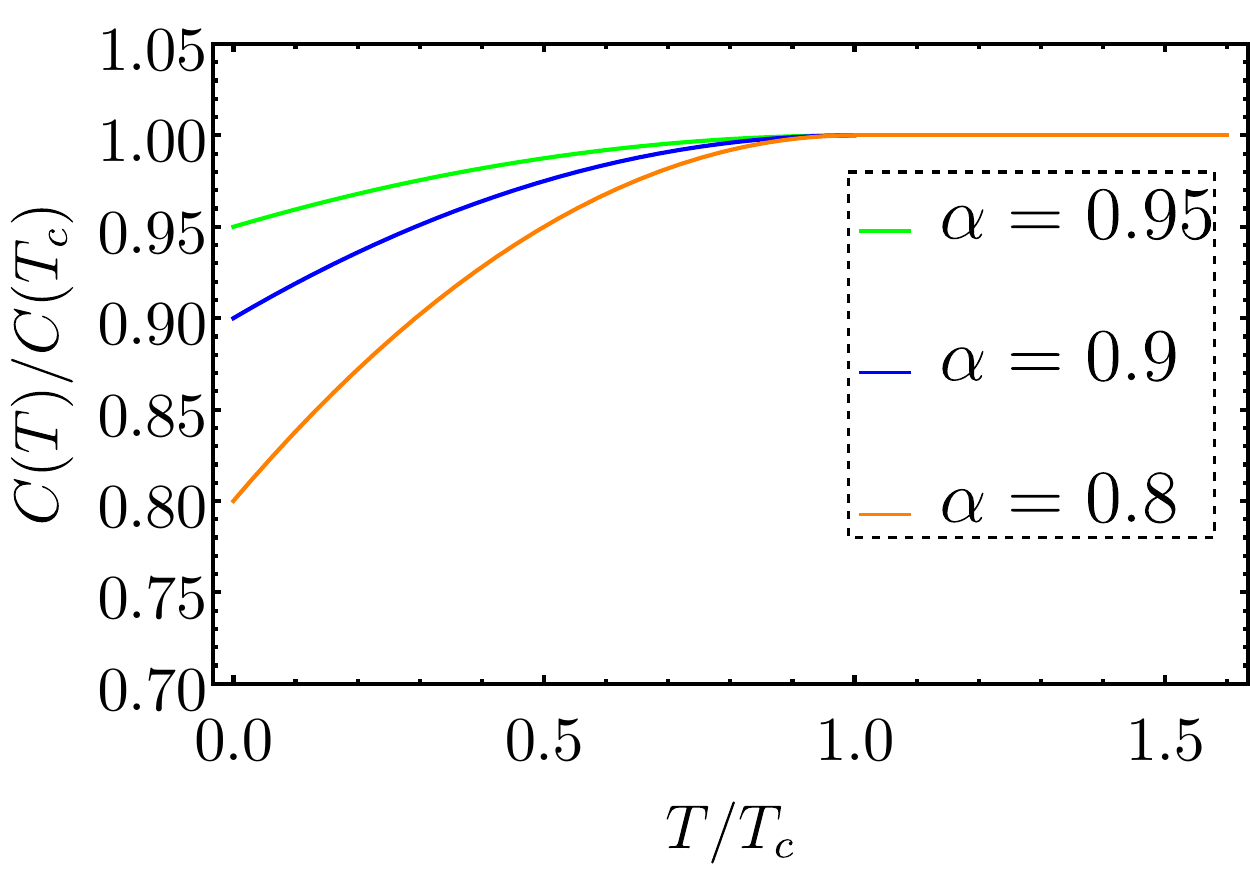}~\label{subfig:alpha}}
\subfigure[]{\includegraphics[width=0.45\linewidth]{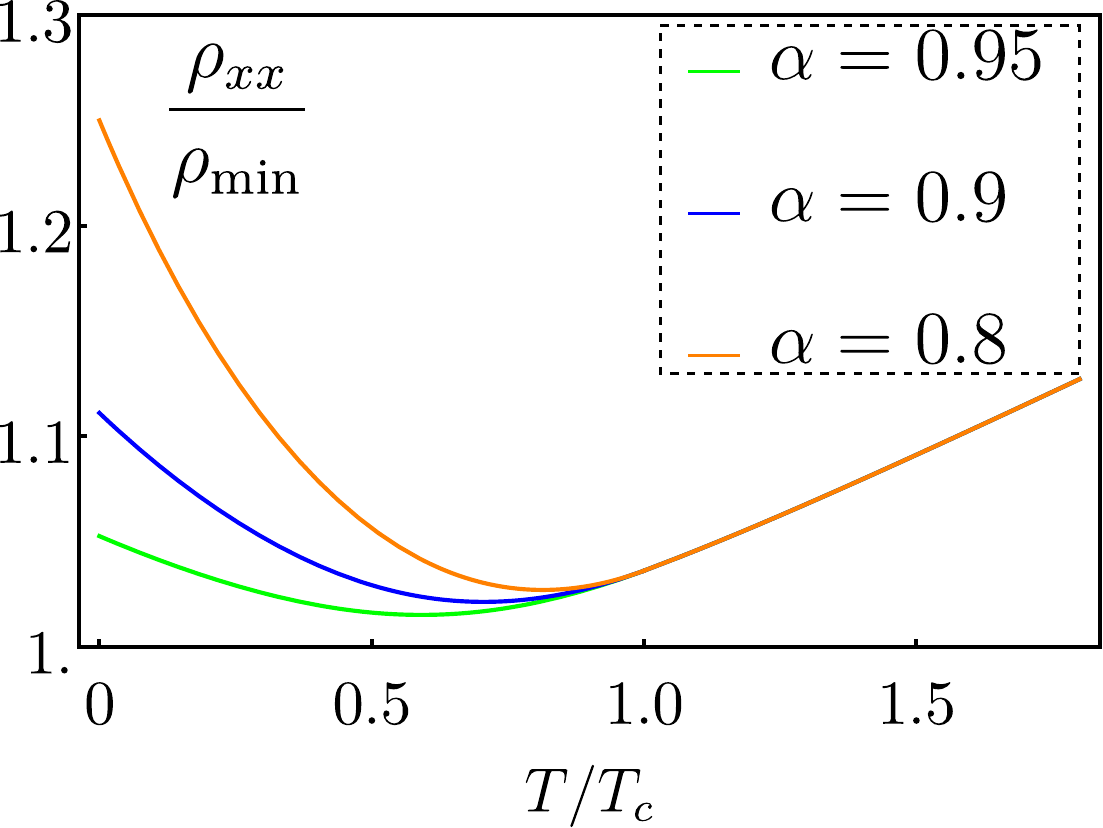}~\label{subfig:Rxx_phenomenology}}
\caption{(a) 
Band curvature of the ordered state $C(T)$ scaled by $C(T_c)$  
in the phenomenological model, Eq.\eqref{eq:alpha-temp}),  for $\alpha = 0.8$ (orange curve), $0.9$ (blue curve), and $0.95$ (green curve), respectively. Here $\alpha$ denotes the ratio $C(T=0)/C(T=T_c)$ 
[see Eq.\eqref{eq:alpha-temp}] \
(b) Plots for resistivity (scaled by the minimum resistivity, which is of the order $\sim \hbar^2 \gamma_0/ n e^2 C(T_c) $) of the phenomenological model for the values of $\alpha$ as in Fig.\ref{subfig:alpha}.
Parameter values used are given in Sec.\ref{sec:phenomenological-model}. 
The resistance initially decreases due to increase in average band curvature, and 
subsequently increases linearly due to phonon scattering.
}~\label{fig:alpha-and-Rxx}
\end{figure}

\section{Phenomenological Picture}\label{sec:phenomenological-model}

Temperature dependence of transport coefficients near pocket-polarization ordering is governed by competition of two distinct effects. Firstly, pocket polarization results in carrier redistribution in $k$ space. As temperature varies, carriers sample different parts of the band, producing temperature-dependent resistivity. Furthermore, because of exchange interactions, the band dispersion and band curvature become sensitive to pocket polarization, providing an additional source of temperature dependence. As our numerical results indicate, the combination of these effects can lead to a negative $d\rho/dT$. Secondly, electron-phonon interaction results in carrier scattering growing at temperatures above the Bloch-Gr\"uneisen temperature, which leads to positive $d\rho/dT$ at high enough temperatures. 

This behavior of the resistivity can be described phenomenologically as follows. At low temperature, the system is in the momentum-polarized ordered
phase, which smoothly turns into a disordered phase
through a continuous phase transition at temperature $T_c$.
The conductivity of the system, 
given by Eq.\eqref{eq:sigma_xx}, depends on the pocket polarization through the quantity $C(T)$, which is the band curvature weighted with the carrier momentum distribution.
At very low temperature, the resistance is primarily affected by the band curvature, as the phonons are not yet thermally activated, and the scattering rate is almost a constant. Above $T_c$, the resistance will primarily increase due to phonon scattering, and the curvature will play a minor role. 
Here, we use a phenomenological temperature dependence of band curvature and scattering rate to demonstrate that at low temperature, the resistance can decrease with rising temperature. Later, in Sec. \ref{sec:microscopic-model}, this behavior will be justified microscopically.

As we will see, the quantity $C(T)$ increases monotonically from $T=0$ to $T=T_c$, 
where the ordered phase 
continuously turns into the disordered phase.
This behavior can be modeled by a 
phenomenological dependence $C(T)$ 
which increases between $T=0$ and $T=T_c$:
\begin{equation}\label{eq:alpha-temp}
    \begin{aligned}
        \frac{C(T)}{C(T_c)} = \begin{cases} 1 - (1-\alpha) \left(1-\frac{T}{T_c} \right)^2 ,& 
        T < T_c \\
&\\
1,& 
T \geq T_c
,
\end{cases}
    \end{aligned}
\end{equation}
with a suitable value $\alpha<1$,
defined as $\alpha = C(T=0)/C(T_c)$. The specific form of Eq.\eqref{eq:alpha-temp} does not matter so long as $C(T)$ grows monotonically. The form of Eq.\eqref{eq:alpha-temp} is chosen so that it mimics the simulation results in the microscopic models. For simplicity, we have set $C(T)/C(T_c)=1$ above $T_c$, because there the temperature dependence of resistance is primarily determined by the phonon scattering. The exact temperature dependence of $C(T)$ obtained microscopically will be established below.

In addition to the band curvature, we must account for the temperature dependence of scattering by phonons and by disorder (defects or impurities),  
\begin{equation}\label{eq:scattering-rate}
	\tau^{-1}(T) =\gamma_\text{ph}(T)+ \gamma_\text{dis}(T).
\end{equation}
At leading order, the disorder scattering rate $\gamma_\text{dis}$ is a temperature-independent constant, $\gamma_0$. The phonon scattering rate $\gamma_\text{ph}$ is given by the Bloch-Gr\"{u}neisen formula \cite{AshcroftMermin,Stormer1990, Hwang-DasSarma2008},
which smoothly interpolates between a power law (quartic in 2D) below the Bloch-Gr\"{u}neisen temperature $T_{\rm BG} = \frac{2 v_s \hbar k_F}{k_B}$, and a linear dependence above this temperature (here $v_s$ is the speed of sound and $\hbar k_F$ is the Fermi momentum). 
In this work, we approximate it with a simple 
phenemenological formula that interpolates between the two regimes, 
$\gamma_\text{ph}(T) = \gamma_1 
\lp(1+(T/T_{\rm BG})^4\rp^{1/4}$, giving 
\begin{equation}\label{eq:phonon-scattering-phenomenology}
        \begin{aligned}
        	\gamma_\text{ph}(T) 
        	 &\approx \begin{cases} \frac{\gamma_1}{4}\left(\frac{T}{T_{\rm BG}}\right)^4, & T \ll T_{\rm BG} \\
        		&\\
        		\gamma_1 \frac{T}{T_{\rm BG}}, &  T \gg T_{\rm BG}
        		.
        	\end{cases}
        \end{aligned}
\end{equation}
Below we set $\gamma_1 = 0.1\gamma_0$, 
so that 
the phonon scattering is relatively weak in the region of phase transition.

In bilayer graphene, the typical value of the Stoner-transition temperature $T_c$, and the Bloch-Gr\"{u}neisen temperature $T_{\rm BG}$ are of the order a few kelvin. So, provided that $T_{\text{BG}}$ is not very small compared to $T_c$, the momentum relazation timescale $\tau(T)$ will not significantly change across the phase transition, and the resistivity will decrease with increasing temperature.
The variation of resistivity with temperature is plotted in Fig.\ref{subfig:Rxx_phenomenology} for several values of $\alpha$ (we have chosen $T_{\rm BG} = 0.8 T_c$ throughout the paper).  When $T_{\rm BG}>T_c$, we find that the resistance continuously decreases with temperature until $T_c$, and increases at higher $T$. When $T_{\rm BG} \lesssim T_c$, the resistance decreases with temperature at $T\le T_{\rm BG}$. Afterwards, between $T_{\rm BG}$ and $T_c$, there will be a competition between the increasing curvature (which suppresses resistance), and the increasing phononic scattering (which enhances resistance). After reaching $T_c$, the resistance will increase with temperature. It is to be noted that this behavior occurs in the metallic regime and at low temperature. The decrease in resistance occurs due to the decreased (average) curvature of the conduction band, which is very different from the phenomena observed in semiconductors, where an initial rise in resistance due to phonon scattering is followed by resistance dropping because the conduction band becomes thermally accessible at higher temperatures.




\section{Microscopic formalism}\label{sec:microscopic-formalism}
Here we introduce a microscopic approach that describes pocket-polarized order 
arising due
to `carrier flocking' in momentum space governed by exchange interaction \cite{ZhiyuIsospinMomentum}. 
Under a generalized Stoner mean field, the instability that leads to pocket asymmetry and pocket polarization is analogous to spin polarization instability in the Stoner magnetism.
We will consider Hamiltonians of the form
\begin{equation}\label{eq:Hamiltonian-form}
    H = \sum_{\vec{k},\sigma} \varepsilon^0_{\vec{k}} c^\dagger_{\vec{k},\sigma} c_{\vec{k},\sigma} + \frac{1}{2} \sum_{\substack{\vec{k}, \vec{k}', \vec{q} \\ \sigma, \sigma'}} V(\vec{q}) c^\dagger_{\vec{k}-\vec{q},\sigma} c^\dagger_{\vec{k}'+\vec{q},\sigma'} c_{\vec{k}',\sigma'} c_{\vec{k},\sigma}.
\end{equation}
Under mean-field theory, we can write the interaction term as a combination of the so-called direct (Hartree) and exchange terms. As long as the average electron density is uniform (e.g. for Bloch waves), the Hartree term is just a constant which would not affect the electronic configurations. 
To the contrary,
 the exchange term favors carriers populating orbitals with equal spin and nearly equal momenta and can drive the instability towards carrier aggregation in
momentum space. After dropping the (constant) Hartree term, we can write the mean-field free energy functional $F=E-TS$ as,
\begin{equation}
    \label{eq:free-energy}
    \begin{aligned}
        & F[f_{\vec{k},\sigma}] = \sum_{\vec{k},\sigma} \varepsilon^0_{\vec{k}} f_{\vec{k},\sigma} - \frac{1}{2} \sum_{\substack{\vec{k}\neq \vec{k}' \\ \sigma}} V(\vec{k}-\vec{k}') f_{\vec{k},\sigma} f_{\vec{k}',\sigma} \\
        &+ k_B T \sum_{\vec{k},\sigma} \left[ f_{\vec{k},\sigma} \ln f_{\vec{k},\sigma} + (1-f_{\vec{k},\sigma}) \ln(1-f_{\vec{k},\sigma}) \right],
    \end{aligned}
\end{equation}
where $f_{\vec{k},\sigma} = \langle c^\dagger_{\vec{k},\sigma} c_{\vec{k},\sigma} \rangle$ and the last term is the entropy contribution written in terms of the occupancy $f_{\vec{k},\sigma}$.

The functional $F[f_{\vec{k},\sigma}]$ defines a variational problem in the functional space $\{f_{\vec{k},\sigma}\}$ in which 
$f_{\vec{k},\sigma}$ are the variational parameters. To tackle this problem, it is convenient to work in the grand canonical ensemble by employing the functional 
\be
\tilde F[f_{\vec{k},\sigma}]=F[f_{\vec{k},\sigma}]-\mu N,
\ee 
where $N=\sum_{\vec k,\sigma}f_{\vec{k},\sigma}$  and $\mu$ is a Lagrange multiplier that fixes the density $N$. To minimize the free energy, we consider the saddle-point conditions $\delta\tilde F/\delta f_{\vec{k},\sigma} =0$, treating $f_{\vec{k},\sigma}$ as independent variables in the functional space. After some algebra the saddle point conditions yield coupled mean-field equations \eqref{eq:mean-field-gen-sol} and \eqref{eq:mean-field-occupation-function}, which describe the critical points of the functional $\tilde F$. 
We use iterations \cite{ThisPaperCodes} to self-consistently solve these equations (for a single spin species) to determine the system ground state.

The unique aspect of the pocket-polarized states is spontaneous symmetry breaking. Normally, in the Landau Fermi-liquid framework, the problem of interacting fermions features a single ground state, representing a Fermi sea defined by the band dispersion $\varepsilon_{\vec k,\sigma}$ renormalized by Fermi-liquid interactions but retaining the symmetry of the free-particle band. The self-consistent equations describing such ground states, derived from the Fermi-liquid theory, are essentially the same as our mean-field equations \eqref{eq:mean-field-gen-sol} and \eqref{eq:mean-field-occupation-function}. 
At not-too-strong interactions $V(\vec k-\vec k')$ or at elevated temperatures, these equations have a self-consistent solution representing a Fermi sea that has the point symmetry group identical to that of the free-particle band dispersion. For a single valley of bilayer graphene, $K$ or $K'$, perturbed by a trigonal warping interaction, it is the discrete symmetry group isomorphic to the $C_{3v}$ group, comprising threefold rotation as well as three valley-preserving operations representing mirror symmetries $\sigma_v$ followed by time reversal.



At a larger interaction strength or at a lower temperature, the self-consistent equations acquire multiple solutions, including three solutions describing broken-symmetry pocket-polarized states forming an orbit of the generalized $C_{3v}$ group and one solution describing a symmetry-unbroken unpolarized state. 
Conceivably, there are two scenarios for these states to emerge from a symmetry-unbroken state upon changing the interaction strength or varying temperature. 
One scenario is when the broken-symmetry solutions, when they first appear, have energies higher than the symmetry-unbroken state. Another scenario is when the broken-symmetry states, when they first appear, are the ground states, whereas the symmetry-unbroken state is a metastable state. In the first case, as the interaction strength increases or temperature decreases, the broken symmetry state emerges abruptly through a type-I transition, whereas in the second case it emerges continuously through a type-II transition. 
In our simulations, described below, the type-II transition scenario is observed upon varying temperature, whereas a type-I transition is seen upon varying carrier density or the interaction strength.

Determining fixed points of the Eqs. \eqref{eq:mean-field-gen-sol} and \eqref{eq:mean-field-occupation-function} corresponding to symmetry-unbroken and symmetry-broken states was done by the method of repeated iterations described below, which was found to converge reliably and rapidly enough.
In our numerical analysis, we found it more convenient to use the canonical ensemble picture with a fixed particle density rather than the grand-canonical ensemble framework. 
To numerically obtain a pocket-unpolarized (symmetry-unbroken) state (metastable below $T_c$) we initiate the iterations with identical carrier distributions in all the pockets. 
This is done by filling up energy states 
from the bottom of each pocket, 
until the total number of electrons reaches the desired value. 
To obtain the pocket-polarized 
ground state with a fixed number of electrons, we initiate the iterations with an out-of-equilibrium state with a particular pocket or two pockets being populated. 
Afterwards, we perform iterations by calculating the renormalized bandstructure with Eq.\eqref{eq:mean-field-gen-sol} and using it to find the new chemical potential, such that the sum of all the 
occupancies $f_{\vec k, \sigma}$ (calculated using Eq.\eqref{eq:mean-field-occupation-function} with a given temperature) equals the desired number of electrons. Plugging these new $f_{\vec k, \sigma}$ in Eq.\eqref{eq:mean-field-gen-sol}, we keep repeating this procedure until a fixed point (symmetry-broken or symmetry-unbroken) is reached. Above $T_c$, the equilibrium distribution becomes pocket-unpolarized (symmetry-unbroken), irrespective of whether the initial distribution of the iteration was pocket polarized or pocket-unpolarized.

\begin{figure*}[t!]
	\subfigure[]{\includegraphics[width=0.31\linewidth]{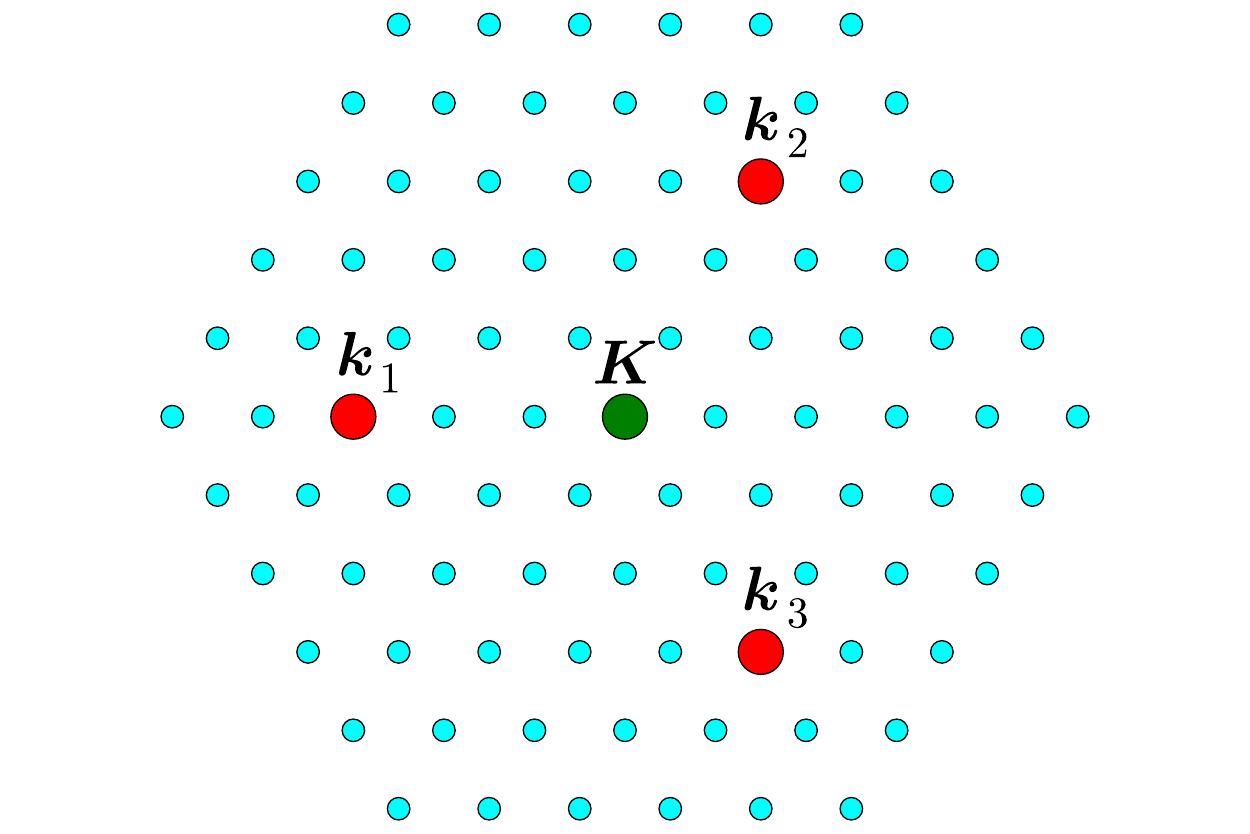}~\label{subfig:triangular_lattice}}
	\subfigure[]{\includegraphics[width=0.315\linewidth]{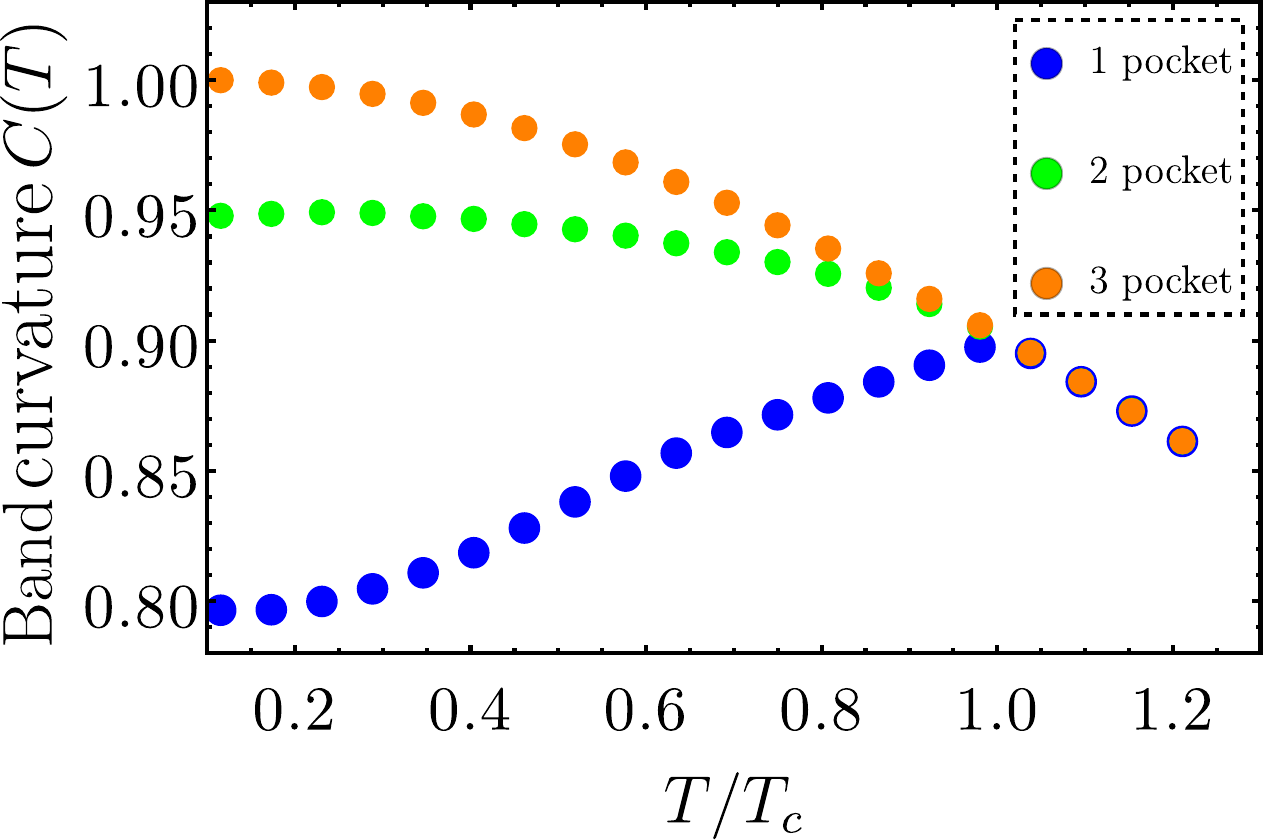}}~\label{subfig:curvature-2d}
	\subfigure[]{\includegraphics[width=0.325\linewidth]{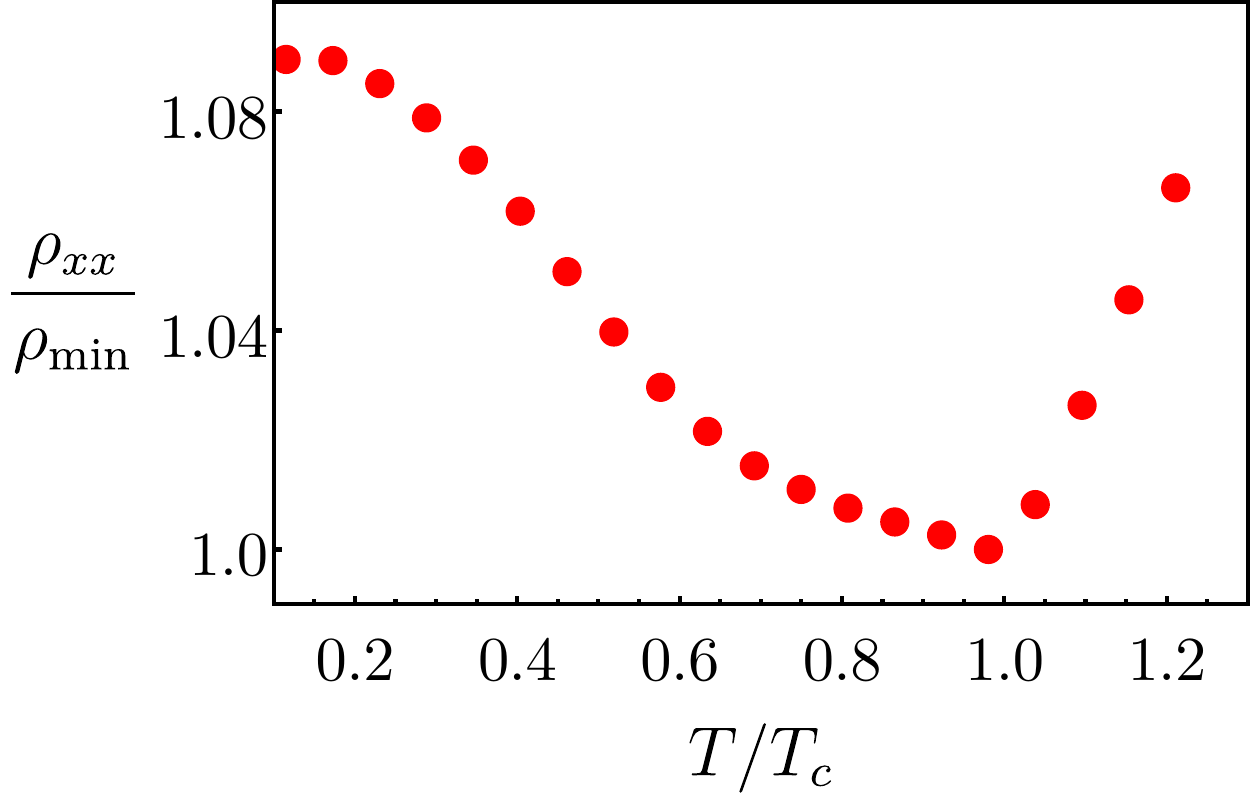}~\label{subfig:resistivity-2d}}
	\caption{(a) A triangular lattice in a hexagonal domain in momentum space, with 6 sites per side of hexagon, illustrating the configuration space used in the 2D model. The pockets are centered around the red points, with distance $k_0 = 3$ from the center (the $\vec{K}$ point, shown in green) of the hexagon. In the actual simulation, we have used 24 sites per side of the hexagon, with $k_0 = 12$. 
		(b) Variation of band curvature obtained numerically for the one-pocket (blue), two-pocket (green) and three-pocket/unpolarized (orange) states of the 2D model. They become identical near $T_c$, until which the curvature of the one-pocket state (ground state) increases as the temperature increases. Parameter values used are given in Sec. \ref{sec:microscopic-model}. 
		(c) 
		Temperature dependence of resistivity for the 1 pocket state. 
		In this model, the resistivity increases very quickly after reaching $T_c$, because of the sharp change in the slope of the band curvature (blue curve in Fig.\ref{fig:2D}(c)). It is really a very quick increase in slope rather than a non-analytic cusp.
	}~\label{fig:2D}
\end{figure*}

\section{
Modeling pocket polarization in 1D and 2D systems}\label{sec:microscopic-model}

Here we show how the behavior outlined in Sec.\ref{sec:microscopic-formalism} emerges in simple models which resemble realistic bandstructures. 
We will start with a one-dimensional double-well band dispersion.
%
%
This simple model captures the pocket-polarized order at low temperature, featuring an ordered phase that continuously turns into a disordered phase at some critical temperature $T_c$. At $T<T_c$, as temperature grows, the degree of pocket polarization diminishes. Simultaneously the occupancy-averaged band curvature $C(T)$ grows and ohmic resistivity exhibits the negative $d\rho/dT$ behavior pictured in Fig.\ref{fig:1D}(d) and \ref{subfig:resistivity-2d}

In this model, the two pockets of the non-interacting dispersion $\varepsilon^0_k$ are situated at some $\pm k_0$, in the region where the band is relatively flat. The form of the interaction $V(\vec k-\vec k')$ is chosen to favor pocket polarization. For that, the inter-pocket interaction must be weaker than the intra-pocket interaction. 
These requirements are met by using the following bandstructure and interaction,
\begin{equation}\label{eq:model1}
    \varepsilon^0_k|_{\text{1D}} 
    = E_0\left(\frac{k^2}{k_0^2}-1\right)^2, \quad 
    V(k-k') = V_0 e^{-\frac{(k-k')^2}{2 q_0^2}}
    ,
\end{equation}
%
where the range of the gaussian interaction $q_0$ is chosen such that the interaction is negligibly small between different pockets, $q_0\ll k_0$. For this model, by minimizing the free energy in  Eq.\eqref{eq:free-energy}, we verify that the system possesses an ordered pocket-polarized  state 
which is thermodynamically stable below $T_c$. As $T$ increases, a pocket-unpolarized state emerges at $T=T_c$ by a type-II transition, whereupon the order parameter vanishes. 
The order parameter describing pocket polarization can be chosen analogously to magnetization, but in terms of pocket populations rather than spins, 
\begin{equation}\label{eq:1D-order-parameter}
	\phi_{1D} = \frac{n_{\text{R}}-n_{\text{L}}}{n_{\text{R}}+n_{\text{L}}},
\end{equation} where $n_L$ and $n_R$ are the total electron numbers in the left and right pockets, respectively. 


The numerical procedure through which an ordered state is obtained follows the description given in previous section.
We initialize the system with a state which has a single pocket populated by all the electrons, and obtain the self-consistent solutions of Eq.\eqref{eq:mean-field-gen-sol} and Eq.\eqref{eq:mean-field-occupation-function} using iterations. 
This selfconsistent  solution is polarized at low temperature, an becomes unpolarized (disordered) at temperatures above some $T_c$. 
We have set $q_0 = 0.2k_0$ and used a mesh of 1001 equidistant points between $k = -1.5 k_0$ to $k=1.5 k_0$, populating the system with $N=190$ electrons so that the low energy excitations fill-up states at the bottom of an empty pocket.
Since the states far away from the bottom of the pocket remain unpopulated and thermally inaccessible at the temperatures we are working with, the boundary conditions on the domain in momentum space do not affect our results, and we use open boundary conditions for convenience. For the simulation, we used dimensionless units $E_0=1$ and the value $V_0=0.004$ for the interaction strength.  

For the values given above, the ordering temperature value was found to be $T_c=0.08$. This energy scale roughly corresponds to the distance between the Fermi level in a filled pocket and the bottom of the empty pocket. 
As discussed in Sec.\ref{sec:transport-anomalies}, there is a competition between thermal excitation to the high energy states of the filled pocket and the states at the bottom of the empty pocket in determining the temperature dependence of band curvature. To thermally excite majority of charge carriers to the bottom of the otherwise empty pocket, we set the density such that the equilibrium distribution at $T\approx 0$ resembles Fig.\ref{fig:1D}(a), where the chemical potential is sufficiently close to the minima of the empty pocket. 
Then, as the temperature is increased, the empty pocket will be accessible to thermal excitation at low temperatures (without exciting very high energy states in the populated pocket), and the system will gradually go to the unpolarized phase. Throughout the process, the average curvature increases as the bottom of the initially unpopulated pocket becomes increasingly populated, and the resistance of the system decreases.

%

Now we describe how 
the band curvature 
was obtained numerically
on a grid.
%
This was done by utilizing a finite-difference method to estimate the second derivative of $\varepsilon_k$, 
\begin{equation}\label{eq:app:1D-lattice-laplacian}
	\Delta^2_k \varepsilon_{k}|_{_{\text{1D}}} = \frac{\varepsilon_{k+\delta k} + \varepsilon_{k-\delta k} - 2\varepsilon_{k}}{(\delta k)^2}.
\end{equation}
The average curvature of the ordered and the (metastable) disordered phases are plotted in Fig.\ref{fig:1D}(c). Unlike the simplistic phenomenological model, where the curvature of the disordered phase was assumed to be a constant, here the curvature of the disordered phase decreases as temperature increases.

We use this microscopic model with Eqs. \eqref{eq:sigma_xx} and \eqref{eq:scattering-rate} to determine the resistivity vs. temperature dependence shown in Fig.\ref{fig:1D}(d). Initially, the resistivity decreases because the average curvature of the ordered state increases as the system smoothly evolves to a
disordered state. Near $T_c$, the average curvature does not increase much, rather it begins to decrease because the system is almost disordered and the high energy states begin to be thermally populated, see Fig.\ref{fig:1D}(c). Eventually, the phonon contribution becomes appreciable at $T> T_{\rm BG}$, which causes an almost linear increase in resistivity for $T > T_c$.

In the second model, we consider a two-dimensional bandstructure with three pockets resembling the pockets induced by trigonal warping\cite{McCann} near $\vec{K}$/$\vec{K'}$ points in bilayer graphene. This model hosts one-pocket and two-pocket polarized phases at low temperature, which continuously turn into a disordered (unpolarized) phase at some $T_c$. 

Here, the relevant order parameter $\vec{D}_k$ is the (dimensionless) dipole moment of electrons in the momentum space, measured from the $\vec{K}$ point,
\begin{equation}
    \vec{D}_k = \frac{\sum_{\vec{k}} (\vec{k} - \vec{K}) f_{\vec{k}}} {\sum_{\vec{k}} k_0 f_{\vec{k}}},
\end{equation}
where $k_0$ is the distance of the center of the pockets from the $\vec{K}$ point. The direction and magnitude of this order parameter vector indicates which pockets have been polarized as well as the nature of the polarization (i.e., one-pocket or two-pocket state).
By construction, the fully polarized one-pocket state has $|\vec{D}_k|=1$, the two-pocket polarized state has $|\vec{D}_k|=0.5$, and the order parameter is 0 for the unpolarized state. As we obtain the equilibrium state at a particular temperature (iterating from a one-pocket or two-pocket polarized initial state), its dipole moment smoothly goes to zero, and that is how the critical temperature was estimated.

We use the following bandstructure and interaction 
\begin{equation}\label{eq:model2}
\begin{aligned}
       & \varepsilon^0_{\vec{k}}|_{\text{2D}} = -E_0 \sum_{i=1}^3 
        e^{-(\vec{k}-\vec{k_i})^2/2 q_1^2},
        \\
       & V(\vec{k}-\vec{k}') = V_0 e^{-(\vec{k}-\vec{k}')^2/2 q_0^2},
\end{aligned}
\end{equation}
where $\vec{k_1} = k_0(-1,0)$ and $\vec{k_{2,3}} = k_0 \left(\frac{1}{2},\pm \frac{\sqrt{3}}{2}\right)$ 
are the centers of the three pockets of radius $q_1$, symmetrically placed on the vertices of an equilateral triangle.

This Hamiltonian was implemented on a triangular lattice in the momentum space (see Fig.\ref{subfig:triangular_lattice}), centered around the $\vec{K}$ point (which is also the center of the $(\vec{k_1},\vec{k_2},\vec{k_3})$ triangle). We use a triangular lattice so that all three pockets are equivalent 
and the discretized problem respects the trigonal symmetry.  
Similar to the previous model, we use open boundary conditions, because the states at the boundary remain unpopulated in the range of electron densities and temperatures we considered.
Parameter values used in our simulation were $E_0 = 1$, $V_0 = 0.01$, $k_0 = 12$, $q_1 = 4$, $q_0 = 6$, and we set $k_B=1$.

The results discussed below were  obtained for a hexagonal domain with trigonal symmetry, with the total of 
$24$ sites per side of the hexagon (see Fig.\ref{subfig:triangular_lattice}).  The electron density was such that in the one-pocket polarized phase at $T\approx 0$, the chemical potential would be very close to the minima of the unoccupied pockets. The array within the hexagon was populated with $N = 82$ electrons so that this condition is satisfied. We find that as the temperature is increased while keeping the electron number fixed, the pocket-polarized phase gradually melts into the disordered phase, and the resistivity decreases. 
In addition to the one-pocket polarized and pocket-unpolarized states, 
a two-pocket polarized state can also be considered. For this particular model, the ground state remains one-pocket polarized at low temperature, and the two-pocket polarized state (as well as the pocket-unpolarized state) is metastable. As temperature is increased, the one-pocket and two-pocket states gradually transform into the unpolarized state, with pocket polarization disappearing at a temperature $T_c$.

In these simulations, the band curvature was evaluated using a discrete lattice-laplacian implemented on the triangular lattice, 
\begin{equation}\label{eq:app:2D-triangular-lattice-laplacian}
	\Delta^2_{\vec{k}} \varepsilon_{\vec{k}}|_{\text{2D}} = \frac{2}{3}\frac{\sum_{\langle\vec{k}'\rangle}\varepsilon_{\vec{k}'}  - 6 \varepsilon_{\vec{k}}}{(\delta k)^2},
\end{equation}
where $\sum_{\langle\vec{k}'\rangle}$ denotes the sum over the six nearest neighbors of $\vec{k}$, with $\delta k$ the distance between nearest neighbors.

The resulting resistivity  vs. temperature dependence is displayed in Fig.\ref{subfig:resistivity-2d}. The resistivity steadily decreases until it reaches $T_{\rm BG}$ because the momentum-averaged curvature decrease  in this regime, and the phonon scattering remains negligible. Afterwards, between $T_{\rm BG}$ and $T_c$, there is a competition between the increasing curvature and the phonon scattering, and after reaching $T_c$, the resistance increases linearly.

\section{Switching induced by electric field} 
\label{sec:switching}

Here we consider pocket polarization switching occurring in the nonlinear current-field response regime. Upon the application of a spatially uniform and time-independent electric field, the whole Fermi sea begins to drift in the direction opposite to the electric field and the relaxation mechanism tries to restore the equilibrium distribution. The net result of this competition is that the Fermi sea is displaced from its equilibrium position in the steady state\cite{AshcroftMermin}. This can be modeled with the Boltzmann transport equation,
\begin{equation}\label{eq:BTE}
	\frac{\partial f}{\partial t} + \frac{d\vec{r}}{dt}\cdot \frac{\partial f}{\partial \vec{r}} + \frac{d\vec{k}}{dt}\cdot \frac{\partial f}{\partial \vec{k}} = -\frac{f - f^0}{\tau},
\end{equation}
where $f$ is the actual nonequilibrium distribution, whereas $f_0$ is the equilibrium distribution defined as a Fermi function of the selfconsistent energy $\varepsilon_k$, a quantity that accounts for the Fermi-liquid interactions with the carrier distribution perturbed by electric field. In the absence of electric field, the selfconsistent energy and the  equilibrium distribution are given by Eqs. (3) and (4). Here, these expressions will be modified as discussed below [Eqs. \eqref{eq:quasiparticle-energy-eigenvalues-shift} and \eqref{eq:steady-state-shifted-distribution}]. 

This approach represents a generalization of the selconsistent relations introduced in the Fermi-liquid theory to an out-of-equilibrium regime. Accordingly, the quasiparticle velocity in the streaming term should be taken as momentum gradient of the selfconsistent energy, $\frac{d\vec{r}}{d t} = \frac{1}{\hbar}\nabla_k \varepsilon_k$. However, for the spatially uniform solutions  considered below the streaming term vanishes. We note that our collision term, which describes relaxation to the selfconsistent equilibrium distribution, agrees with the requirements of the Fermi-liquid theory 
\cite{PinesNozieresQuantumLiquids, PitaevskiiLifshitzKinetics}.
Here, for simplicity, we employ the relaxation time approximation, and assume that the relaxation timescale is momentum independent.

In the steady state, $\frac{\partial f}{\partial t} = 0$, and $\frac{\partial f}{\partial \vec{r}}=0$ due to spatial homogeneity. Also, $\frac{d\vec{k}}{dt} =-\frac{e \vec{E}}{\hbar}$ from the semiclassical equations of motion, where the charge of an electron is $-e$. Then, the equation reduces to
\begin{equation}\label{eq:BTE-reduced}
	f(\vec{k}) - \frac{e\tau}{\hbar}\vec{E}\cdot \frac{\partial f(\vec{k})}{\partial \vec{k}} = f^0(\vec{k})
	.
\end{equation}
This equation has a formal solution,
\begin{equation}\label{eq:BTE-reduced-gen-sol}
	f(\vec{k}) = \frac{1}{1 - \frac{e\tau}{\hbar}\vec{E}\cdot \frac{\partial}{\partial \vec{k}}}\cdot f^0(\vec{k}).
\end{equation}
To evaluate it, either the operator $\frac{1}{1 - \frac{e\tau}{\hbar}\vec{E}\cdot \frac{\partial}{\partial \vec{k}}}$ can be written as a Taylor series in $\vec{E}\cdot\frac{\partial}{\partial \vec{k}}$, or more formally, it can be evaluated using Fourier transforms, giving
\be \label{eq:BTW-kernel-solution}
f(\vec k)=\int_0^\infty ds e^{-s} f^0(\vec k+se\vec E\tau/\hbar)
,
\ee
where $s$ is an auxiliary integration variable.

At linear order in the electric field, the solution takes the form 
\begin{equation}\label{eq:BTE_sol}
	f(\vec{k}) \approx f^0(\vec{k}) + \frac{e\tau}{\hbar}\vec{E}\cdot \frac{\partial f^0(\vec{k})}{\partial \vec{k}} \approx f^0 \left(\vec{k} + e \vec{E} \tau/\hbar \right),
\end{equation}
that is, the steady state Fermi distribution $f$ will be a displaced version of the equilibrium Fermi distribution $f^0$, as described above.
If we take a pocket-polarized state, apply an electric field in the appropriate direction, and keep increasing its magnitude, at some point the displaced Fermi sea may 
abruptly switch, resulting in all the electrons shifting to another pocket. Since a finite electric field is required for the switching, the initial state cannot be restored by decreasing the magnitude of the electric field, or by the application of a small electric field in the opposite direction. Experimentally, such
memory effects and history-dependent behavior in the transport properties may be observed by turning on a strong in-plane electric field, or by running high currents. These results, obtained here for a simple model, are expected to describe transport  in a bilayer (or multi-layer) graphene or transition metal dichalcogenides (TMD) sample that hosts multiple Fermi pockets induced by trigonal warping\cite{Seiler, Li1, ZhiyuIsospinMomentum}. The switching behavior, which is a consequence of pocket polarization assisted by an applied $E$ field, may be used as an experimental probe for delineating pocket polarization from alternative mechanisms  of negative $d\rho/dT$.

To model the switching effect, we modify the self-consistent equations \eqref{eq:mean-field-gen-sol} and \eqref{eq:mean-field-occupation-function} as follows. The exchange energy will depend on 
the field-induced steady-state distribution, modeled as the shifted equilibrium distribution. 
as discussed above. Namely, the coupled mean-field equations \eqref{eq:mean-field-gen-sol} and \eqref{eq:mean-field-occupation-function} must be replaced with the three equations
\begin{subequations}
	\label{eq:E_field_shifted_equations}
	\begin{align}
		\begin{split}
			\varepsilon_{\vec{k},\sigma} = \varepsilon^0_{\vec{k}} - \sum_{\vec{k}'\neq \vec{k}} V(\vec{k}-\vec{k}') f_{\vec{k}',\sigma},
		\end{split}
		\label{eq:quasiparticle-energy-eigenvalues-shift}
		\\
		\begin{split}
			f_{\vec{k},\sigma} = f^0_{\vec{k} + e \vec{E} \tau/\hbar,\sigma},
			\quad
			f^0_{\vec{k},\sigma} = \frac{1}{e^{\beta (\varepsilon_{\vec{k},\sigma} - \mu)}+1}.
		\end{split}
		\label{eq:steady-state-shifted-distribution}	
\end{align}
\end{subequations}
To numerically solve these equations, we use the 1D model (Eq.\eqref{eq:model1}) with $E_0 = 1$, $k_0 = 1$, $V_0 = 0.01$, $q_0 = 0.2$, $k_B T/E_0 = 0.2$. We 
use the mesh of 2001 equidistant points between $k = -3 k_0$ to $k = 3 k_0$, and populating it with $N = 200$ electrons. We initiate the iterations with an equilibrium distribution polarized in the right (red curve in Fig.\ref{fig:E_field_switching}) as well as the left pocket (blue curve in Fig.\ref{fig:E_field_switching}), and in each case obtain a field-induced steady-state distribution.

To clarify the underlying physics, 
consider the evolution of the right pocket-polarized initial state. Under the application of a small positive electric field, the steady state distribution remains polarized at the same pocket, but as the electric field is increased, the distribution immediately switches to the left pocket, and the pocket polarization $\phi_{1D} = (n_R - n_L)/(n_R + n_L)$ discontinuously jumps from $+1$ to $-1$. The effect has been illustrated in Fig.\ref{fig:E_field_switching}, where the switching occurs at $E \approx 0.32 \hbar k_0/e \tau$ for the parameters mentioned in the previous paragraph. If we had swept the electric field in the opposite direction from the beginning (for the same right pocket-polarized initial state), then the steady state distribution would shift to the direction opposite to the left pocket. In this case, the system will not exhibit any switching behavior, but the pocket polarization $\phi_{1D}$ will gradually decrease in magnitude. Due to the displacement of the Fermi sea, the exchange energy at the minima of the filled pocket would be smaller compared to the initial state. Consequently, energy difference between the minima of the filled and the empty pockets would be smaller, and the initially empty pocket will gain some electrons due to thermal excitations. Therefore, the sign of $\phi_{1D}$ would remain the same, but it would decrease in magnitude, as observed in Fig.\ref{fig:E_field_switching}. The analogous and opposite effect happens with the left pocket polarized initial state.

\section{Discussion 
}\label{sec:discussion}

The transport anomalies considered here, originating from pocket polarization, involve a negative temperature dependence of resistivity predicted in the ohmic regime and polarization switching in a non-ohmic regime.
We first restate the reasons for negative $d\rho/d T$ being a robust and generic property of the momentum-polarized order.
At low temperatures, the electrons are predominantly scattered by disorder as the phonons are not yet thermally activated. Consequently, the resistivity of a metallic system is primarily determined by the average band curvature. Using microscopic models that mimic the pockets induced by trigonal warping in bilayer graphene and TMD, we demonstrated that the momentum-polarized ordered phase samples a greater number of states with relatively less curvature compared to its disordered counterpart, implying that the resistivity of the ordered phase will decrease with rising temperature, until the order undergoes melting. Subsequently, phonon scattering will reinstate the linear dependence of resistivity on temperature, just like conventional metals at room temperature.

Simultaneously, transport in the momentum-polarized phase features a dependence on history that leads to polarization switching.
Namely, a particular momentum-polarized state can switch to a different momentum-polarized state when subjected to a strong electric field. 
This behavior is expented to give rise to hysteretic $I$-$V$ characteristics 
that may be utilized to experimentally distinguish the mechanism of negative $d\rho/dT$ described above from other mechanisms.

A similar behavior in resistivity 
may be considered for the `valley-polarized', and `spin-polarized' ordered phases at very 
low carrier densities. To understand the microscopic picture, consider the valley-polarized phase and, for the time being, ignore  pocket polarization. At low temperature, while only one valley is polarized, the resistance will gradually increase with rising temperature, just like the behavior observed in ordinary metals. As $T$ increases, at some point the bottom of the unpopulated valley will begin to be populated by thermally excited carriers. These carriers will sample  the bottom part of the band where the curvature is high.  The higher average curvature of the populated part of the band will bring the  resistance down. Eventually, the thermal excitations will suppress valley-polarized order, and the resistance will begin to increase again. Therefore, as $T$ grows, resistance will rise, then drop and then rise again. 

A similar $T$ dependence is expected to occur in the spin-polarized phase. Pocket polarization, if present, will make the behavior more complicated. However, it is unlikely to eliminate the nonmonotonic $T$ dependence.
However, for the scenario discussed above to be applicable at much higher density (where spin or valley-polarized phases are usually not observed), the bandstructure 
must be very different from standard bilayer graphene bandstructure near charge neutrality. Experimentally, 
the phases polarized in spin or valley have only been stabilized at low electronic density (close to charge neutrality). Negative $d\rho/dT$ may be observed if the ordered state can be stabilized at a high enough electronic density so that it will sample an appreciable number of states with negative band curvature from the high energy necks linking the $K$ or $K'$ valleys, whereas the corresponding high-temperature state will sample a relatively greater number of states from the valley bottom, which has a positive band curvature. 
A bandstructure of this type is featured by biased antimonene, whose conduction band has several almost degenerate pockets \cite{AntimoneneWang2015, AntimoneneRoldan2017} which at a low temperature may host pocket-polarized phases.
Further research is required to explore the resistive behavior of these systems. 
The longitudinal resistivity decreasing with temperature, combined with the switching behavior can serve as a transport signature to experimentally identify 
momentum-polarized ordered phases. 

\section*{Acknowledgements}
We are grateful to Andrea Young for sharing unpublished data and useful discussions, and to Dmitri Maslov for enlightning comments on the origin of a negative $d\rho/dT$ observed in silicon MOSFETs. This work was supported by the Science and Technology Center for Integrated Quantum Materials, National Science Foundation Grant No. DMR1231319. 

\onecolumngrid
\begin{center}
	\rule{0.6 \linewidth}{0.4pt}
\end{center}

\bibliographystyle{plain}
\end{document}